\documentclass[preprint]{aastex62}

\submitjournal{ApJ Letters}

\begin{document}

\title{Satellite Galaxies in the Illustris-1 Simulation: Poor Tracers of the
Mass Distribution}

\correspondingauthor{Tereasa G. Brainerd}
\email{brainerd@bu.edu}

\author[0000-0001-7917-7623]{Tereasa G. Brainerd}
\affil{Boston University \\
Department of Astronomy \\
725 Commonwealth Avenue \\
Boston, MA 02215 USA}

\begin{abstract}

Number density profiles are computed for the satellites of relatively isolated host 
galaxies in the Illustris-1 simulation.
The mean total mass density of the hosts
is well-fitted by an NFW profile.
The number density profile for the complete satellite sample is inconsistent with
NFW and, 
on scales $\lesssim 0.5~r_{200}$, the satellites do not trace the hosts' mass.
This differs 
substantially 
from previous results from semi-analytic
galaxy formation models.
The shape of the satellite number density profile depends
on the luminosities of 
the hosts and the satellites, and on the host virial mass.
The number density profile
for the faintest satellites 
is well-fitted by an NFW profile,
but the concentration is much less than the mean
host mass density.  The number density profile for
the brightest satellites 
exhibits a steep increase in
slope for host-satellite distances $\lesssim 0.1~r_{200}$, in qualitative 
agreement with recent observational studies that
find a steep increase in the satellite number density at small host-satellite
distances.
On scales $\gtrsim 0.1~r_{200}$ the satellites of the faintest hosts
trace the host mass reasonably well.
On scales $\lesssim 0.4~r_{200}$, the satellites of the brightest hosts
do not trace the host mass and
the satellite number density increases steeply for host-satellite distances
$\lesssim 0.1~r_{200}$.
The discrepancy between the satellite number density profile and the
host mass density is most pronounced for the most massive systems,
with the satellite number density falling far below that of the mass density
on scales $\lesssim 0.5~r_{200}$.
\end{abstract}



\section{Introduction} 

In 
$\Lambda$ Cold Dark
Matter ($\Lambda$CDM), dark matter halos follow a ``universal''
shape that is often parameterized as the Navarro, Frenk 
\& White (NFW; Navarro et al.\ 1996, 1997) profile,
\begin{equation}
\rho(r) = \frac{\delta_c \rho_c}{(r/r_s) (1+r/r_s)^2},
\end{equation}
where $\rho_c$ is the critical density for closure of the universe and
$r_s \equiv r_{200}/c$ is the scale radius. The virial radius, $r_{200}$, is 
the radius for which the mean
interior halo mass density is $200 \rho_c$. 
The concentration, $c$, is related to the characteristic overdensity,
$\delta$, 
through: 
\begin{equation}
\delta_c = \frac{200}{3} \frac{c^3}{\ln(1+c) - c/(1+c)}.
\end{equation}
An important 
test of $\Lambda$CDM is the degree to which the dark matter
halos of bright galaxies follow an NFW-like
density profile.  
In order to directly probe the dark matter distribution around
bright galaxies, a luminous tracer of the mass is
necessary. In practice, small, faint satellite galaxies, in orbit around
large, bright ``host'' galaxies, could be fair tracers of the mass,
and it is the degree to which these objects trace the mass surrounding
their hosts that is the subject of this investigation.

The degree to which observed satellite galaxies trace their
hosts' mass is not settled.
Budzynski et al.\ (2012) 
found that the number density profile for satellites in groups and
clusters was well-fitted by an NFW profile, but the concentration of the
satellite distribution was a factor $\sim 2$ lower than expected for
the dark matter.  
In another cluster study,
Wang et al.\ (2018) compared the mass density profile
obtained from weak lensing 
to the satellite distribution and concluded that 
both profiles were 
fitted well by NFW, with the satellites tracing the mass.
Nierenberg et al.\ (2012) concluded that satellites of massive
galaxies ($M_\ast > 3\times 10^{10} M_\odot$) traced the mass
distribution well.  Similarly, Guo et al.\ (2013) concluded that,
with the exception of small differences at small radii and low luminosity,
the satellite number density profile for isolated host galaxies was a good
tracer of the mass.
Wang et al.\ (2014)
found that the satellites of relatively
isolated host galaxies with stellar masses $M_\ast > 10^{11} M_\odot$
had a spatial distribution that was less concentrated than would be expected
for the hosts' dark matter halos, while the satellites of less massive hosts
had steeper density profiles that agreed well with the expected dark matter 
distribution.  
Tal et al.\ (2012) found that, 
on scales $\gtrsim 270$~kpc,
the number density profile for the satellites of Luminous
Red Galaxies (LRGs) was well-fitted by NFW.
On small scales, however, Tal et al.\ (2012) found a distinct upturn in the 
satellite number density profile that is inconsistent with NFW.  
Further, Watson et al.\ (2012) and
Piscionere et al.\ (2015) concluded that the small scale
($\lesssim 40h^{-1}$~kpc) clustering of bright satellite galaxies 
($M_r < -20$) was consistent with a number density profile that
was much steeper than NFW.

In the theoretical regime, the degree to which satellites trace 
their hosts' mass is also not settled.
Gao et al.\ (2004) combined a semi-analytic 
galaxy formation model (``SAM'') with high-resolution
$N$-body simulations of the formation of massive galaxy clusters and
concluded that luminous galaxies trace the cluster mass well.
From numerical hydrodynamics simulations,
Nagai \& Kravtsov (2005) concluded that the concentration of the satellite 
distribution inside clusters was less than that of the
dark matter.
Sales et al.\ (2007; hereafter SNLWC) computed the number density profile for satellite
galaxies in the Millennium simulation (MS; Springel et al.\ 2005)  using
a luminous galaxy catalog obtained from a SAM.
SNLWC focused 
on relatively isolated host galaxies at $z = 0$, the vast majority of which were 
central galaxies within the surrounding dark matter distribution. 
SNLWC concluded that: [i] the satellite number
density profile was well-fitted by NFW,
[ii] the shape of the satellite
number density profile did not depend strongly on the luminosity of the
host galaxy or its halo virial mass, and [iii] the distribution of the 
satellite galaxies was similar to that of the dark matter, but slightly
less concentrated.  
Wang et al.\ (2014) computed number density profiles for satellite
galaxies in the MS and the Millennium-II simulations (MS-II; Boylan-Kolchin
et al.\ 2009), where the luminous galaxy catalog was obtained from
a SAM.  Wang et al.\ (2014) found
that the satellites of relatively isolated host galaxies 
traced the mass distribution of the hosts' halos well and there was little
dependence of the satellite number density profile on the physical properties
of the hosts. 
Ye et al.\ (2017) explored the spatial distribution of
satellites with large stellar masses ($\ge 10^{9} h^{-1} M_\odot$) in
the hydrodynamical Illustris-1 simulation (Vogelsberger et al.\ 2014a; Nelson
et al.\ 2015).  At 
$z = 0$, Ye et al.\ found that for halos 
with virial masses $10^{12} h^{-1} M_\odot < M_{200} < 10^{14} h^{-1}
M_\odot$, the dark matter mass within $\lesssim 0.4 r_{200}$ was better
traced by satellites with a high satellite-to-host mass ratio than it
was by satellites with a low satellite-to-host mass ratio.
\'Ag\'ustsson \& Brainerd (2018) investigated the spatial distribution
of satellite galaxies in the MS that were selected using redshift space
criteria and  
found that
the satellite distribution was a good tracer of the
halos of red hosts,
but the satellite distribution around blue hosts
was roughly twice as concentrated as the hosts' halos.


Here, 
the hydrodynamical Illustris-1 simulation
is used to obtain the 
number density profiles for the satellites of relatively
isolated host galaxies.  The
key questions that are addressed are: [1] the degree to which satellite
galaxies trace the mass distribution, 
and [2] how the shape of the satellite number density profile
compares to the shape
obtained previously from SAMs.
The paper is organized as follows.  The host and satellite selection criteria,
and the properties of the sample are presented in \S2.  Satellite number
density profiles and host galaxy mass density profiles are presented in 
\S3.  A summary and discussion of the results is presented in \S4.

\section{Host-Satellite Sample} 

Illustris-1 
followed the growth of structure in a $\Lambda$CDM universe using
$\Omega_m = 0.2726$, $\Omega_\Lambda = 0.7274$, $\Omega_b = 0.0456$,
$\sigma_8 = 0.809$, $n_s = 0.963$, and $H_0 = 70.4$~km~s$^{-1}$~Mpc$^{-1}$.
The simulation volume was a cubical box with periodic boundary
conditions and comoving sidelength $L = 106.5$~Mpc.  A total of
$1820^3$ dark matter particles of mass $6.3\times 10^6 M_\odot$ and
$1820^3$ hydro cells with initial baryonic mass resolution of 
$1.26\times 10^6 M_\odot$ were used.  Here, 
only the $z = 0$ timestep is used, for which the force 
softening length is $\epsilon_{dm} = 710$~pc, the smallest hydrodynamical gas cells
are 48~pc in extent and there are $\sim 40,000$ luminous galaxies.

Host galaxies were obtained using the criteria adopted by SNLWC.
Host galaxies have absolute magnitudes $M_r < -20.5$
and, within a radius of $1h^{-1}$~Mpc centered on the host, are
surrounded {\it only} by companions at least two magnitudes fainter than
the host.  Host
galaxies must also have at least one companion (i.e., a satellite galaxy)
within $r_{200}$.  In addition, Illustris-1 hosts were 
required to be located at the centers of their friends-of-friends halos.  
In SNLWC the satellite galaxies were restricted to 
objects with 
$M_r <  -17$ due to the MS resolution limit. 
In Illustris-1 the resolution is such that satellites as
faint as $M_r = -14.5$ can be resolved (see, e.g., 
Vogelsberger et al.\ 2014b) and, so, these additional
faint satellites are also included here.
Imposing these criteria  
results in a total of 1,025 Illustris-1 host
galaxies with at least one satellite within $r_{200}$.  The total number of satellites
within $r_{200}$ of all host galaxies is 4,546. 
Within $r_{200}$, the number of satellites 
for a given host ranges from 1 to 306, with a median of 2.
The median host virial mass is $M_{200}^{\rm host} = 10^{12} M_\odot$, the median host
absolute magnitude is $M_r^{\rm host} = -21.8$, and the median host stellar mass is 
$M_\ast^{\rm host} = 3.2\times 10^{10} M_\sun$.  The median host stellar mass is $\sim 2,100$
times larger than the median satellite stellar masses ($M_\ast^{\rm sat} = 1.5\times 10^7 M_\odot$)
and the median host-satellite luminosity ratio is $\sim 2,500$.
Because the host galaxies were selected using the same criteria,
the Illustris-1 host virial mass distribution is similar to that
in SNLWC (see their Figure~2).
Figure~1 summarizes various properties of the sample.

\begin{figure}[t!]
\plotone{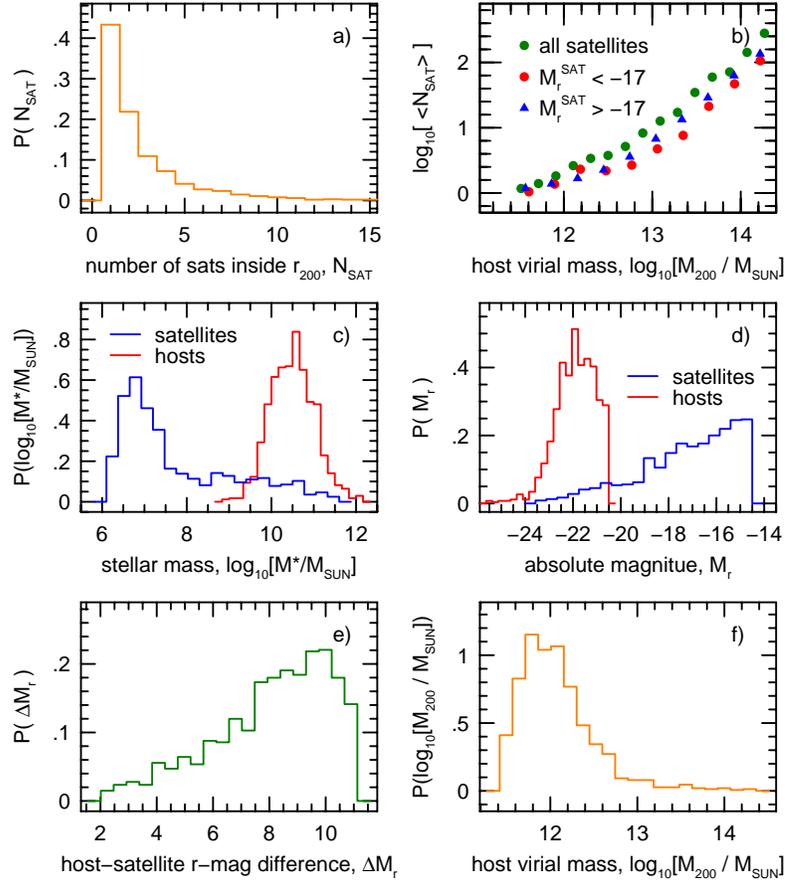}
\vskip -6.0cm
\caption{Properties of the sample: a) distribution of
the number of satellites
within $r_{200}$, b) number of satellites within $r_{200}$ as 
a function of host virial mass, 
c) stellar mass distributions for hosts and satellites,
d) absolute $r$-band magnitude distributions for hosts and satellites, 
e) distribution
of host-satellite 
$r$-band absolute magnitude differences, f) host virial mass distribution.}
\label{fig:sample}
\end{figure}

\section{Density Profiles}

Density profiles for the satellite distribution (i.e., the mean
satellite number density) and the mass surrounding the host galaxies, including both
baryonic and dark matter, were computed. Following SNLWC, the
density profiles were
normalized by their values at $x \equiv r/r_{200} = 1$.  Error bars 
were computed using boostrap resampling and are omitted from figures when
they are comparable to or smaller than the sizes of the data points.  The top panel of
Figure~2 shows the normalized
density profiles for the distribution of all satellite galaxies and
the mean mass density for the hosts, where the mean mass density is
computed as an unweighted mean over all hosts.  Also shown 
is the best-fitting NFW profile for the unweighted mean host mass density.  
The unweighted mean mass density of the hosts is well-fitted by an NFW profile, and the 
concentration of the best fit is
$c = 11.9$.  
On scales $\lesssim 0.4~ r_{200}$, the satellite distribution does not trace the
hosts' mass.  In addition, due to the inflection
around $x = 0.2~ r_{200}$, the mean satellite number density cannot be fitted by
an NFW profile.

The middle panel of Figure~2 shows the density profiles for the brightest
satellites ($M_r < -17$), corresponding to the satellites in SNLWC,
together with the density profiles for the
faint satellites that are resolved in Illustris-1 but not 
the MS.  
Since the median absolute magnitude of the Illustris-1 satellites is 
$M_r^{\rm med} = -16.8$, division of the satellites into those with
$M_r < -17$ and those with $M_r > -17$ effectively splits the sample in half.
From the middle panel of Figure~2, 
the Illustris-1 satellites that are comparable in luminosity to 
SNLWC's satellites have a number density profile that cannot be
fitted by an NFW profile, again due to the steep upturn at 
small host-satellite separations.  In contrast, the number density profile for
Illustris-1 satellites with
$M_r > -17$ is well-fitted by an NFW profile. However, the
concentration of the best-fitting NFW profile for the number density of the faintest
satellites, $c = 1.8$, is a factor of 6.5 less than that of the unweighted mean host
mass density.  
That is, 
the faintest satellites, while following an NFW profile, have a spatial distribution
that is significantly less concentrated than the underlying mass.

\begin{figure}[t!]
\plotone{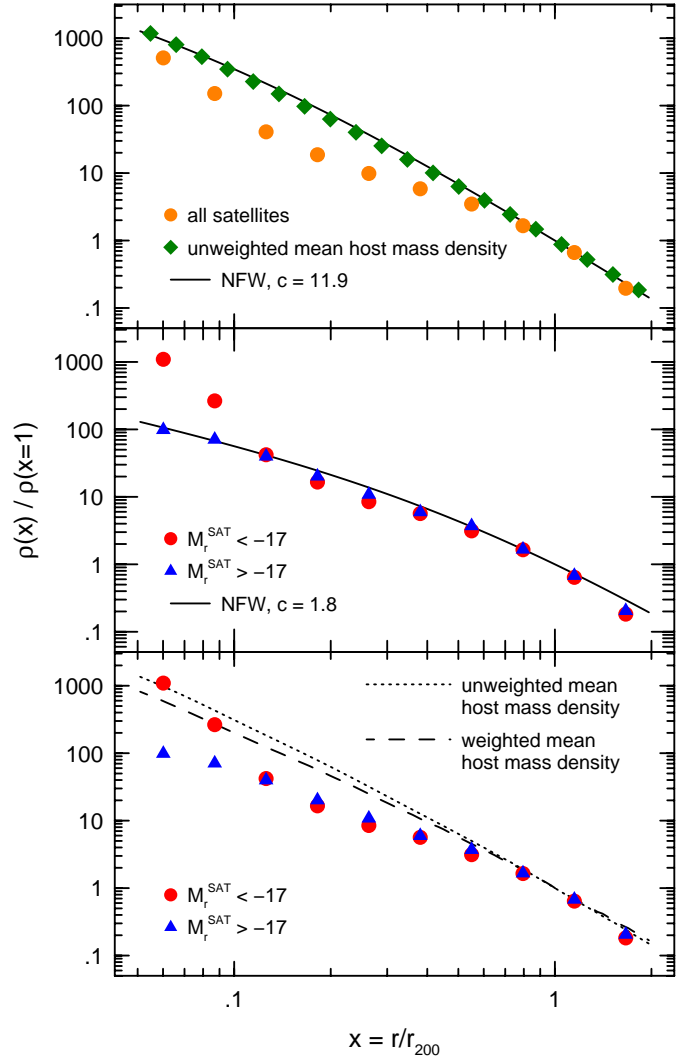}
\vskip -2.5cm
\caption{Normalized satellite number density profiles and host mass density profiles.
{\it Top:} all satellites (orange circles), host mass density computed
as an unweighted mean
(green diamonds),
and best-fitting NFW profile for the host mass density (solid black line).
{\it Middle:} satellites with $M_r$ comparable to the satellites in SNLWC
(red circles), satellites with $M_r$ fainter than the satellites in 
SNLWC (blue triangles), and best-fitting NFW profile for the faint
satellites.
{\it Bottom:} 
Faint and bright satellites from the middle panel, and
host mass density computed as both an unweighted mean and
a weighted mean where the weights correspond to the number of
satellites per host.
}
\label{fig:profiles1}
\end{figure}

Since the number of satellites per host increases 
with host virial mass, more massive hosts
contribute a greater number of satellites to the satellite number density
profile than do less massive hosts.  To assess the effects of this on
the degree to which the satellites trace the host mass density, the 
bottom panel of Figure~2 shows the satellite number density profiles 
of the faint and bright satellites (i.e., from the middle panel of 
Figure~2), along with the mean host mass density computed in two ways:
an unweighted mean and a weighted mean in which the 
weights correspond to the number of satellites within $r_{200}$.
From the bottom panel of Figure~2, the mean 
host mass density, weighted by the number of satellites for each host,
results in a less concentrated mean host mass density profile
($c = 7.3$ for the weighted mean vs.\
$c = 11.9$ for the unweighted mean).
This does not, however,
significantly affect the overall disagreement between
the satellite number density
profile and the mean host mass density profile.

Figure~3 shows the dependence of the satellite number density profiles
on host luminosity, with the results for the brightest 1/3 of the hosts 
($M_r < -22.5$) shown in the top panel and the results
for the faintest 1/3 of the hosts ($M_r > -21.5$) shown in the bottom panel.
As in Figure~2, the satellites are split into ``faint'' and ``bright''
samples, and both uweighted and weighted mean host mass density
profiles are shown.  Since the number of satellites per host varies
little for the hosts in the top and bottom panels of Figure~2,
the weighted and unweighted mean host
mass densities in these panels are
essentially identical.  In the case of the most luminous hosts (for
which the number of satellites per host varies signficantly), 
the weighted mean host mass density is somewhat less concentrated
than the unweighted mean.
 
From Figure~3, it is clear that in no case do the satellites trace the 
host mass over all scales that are resolved by the simulation.  
(Note: the smallest host-satellite separation shown in Figure~3,
$r/r_{200} \sim 0.06$, corresponds to $\sim 10 \epsilon_{dm}$
for the host with the smallest virial radius, $r_{200} = 123$~kpc.)
The shape of the satellite number density profile depends on
host luminosity, and it deviates most significantly from the host
mass density in the case of the most luminous hosts.
In the case of the faintest hosts, the satellites trace the host mass reasonably
well on scales $\gtrsim 0.1~r_{200}$. {The satellite number density 
profile of the faintest hosts 
shows little dependence on satellite luminosity. For
the brighter host galaxies, however,
the number density of the bright satellites is 
exceeds that of the faint satellites on scales 
$\lesssim 0.1 r_{200}$.
}

\begin{figure}[t!]
\plotone{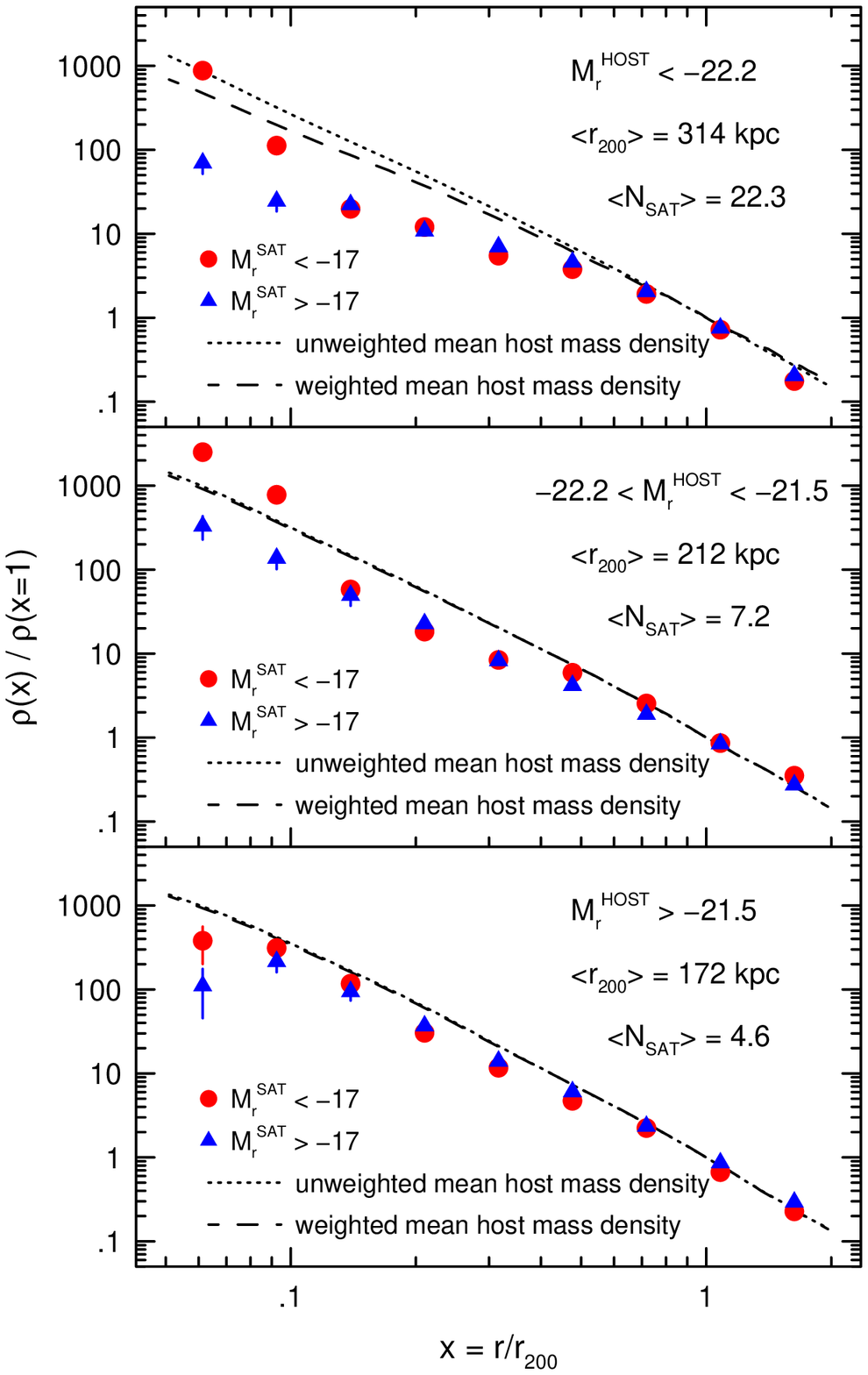}
\vskip -2.5cm
\caption{Normalized satellite number density profiles 
and host mean mass density profiles as a function of host luminosity.
Satellites are divided in luminosity as in Figure~2.
Unweighted means and means weighted 
by the number of satellites per host are shown
for the host mass density.
{\it Top:} brightest hosts ($M_r < -22.2$; 355 hosts).
{\it Middle:} hosts with $-22.2 \le M_r \le -21.5$ (301 hosts).
{\it Bottom:} faintest hosts ($M_r > -21.5$; 369 hosts).
}
\label{fig:profiles2}
\end{figure}

Figure~4 shows the dependence of the satellite number density profile on
host virial mass.  The top panel shows the most massive 9\% of the systems, for
which the deviation of the satellite number density profile from the host
mass density is particularly pronounced.  The other panels show lower
mass hosts, split into two samples of roughly equal size.
As in Figure~3, the host mass density profiles were computed as
both unweighted means and means weighted by the number of satellites per
host.  It is only in the case of the most massive hosts that weighting the
mean host mass density by the number of satellites has any
noticeable effect on the resulting density profile, and it does not significantly
affect the degree to which the satellites trace the host mass density.
The number density profiles for the
satellites of the most massive systems show little dependence on satellite
luminosity.  In the case of the less massive hosts, the number density
of the bright satellites exceeds that of the faint satellites
on scales $\lesssim 0.1 r_{200}$.

\begin{figure}[t!]
\plotone{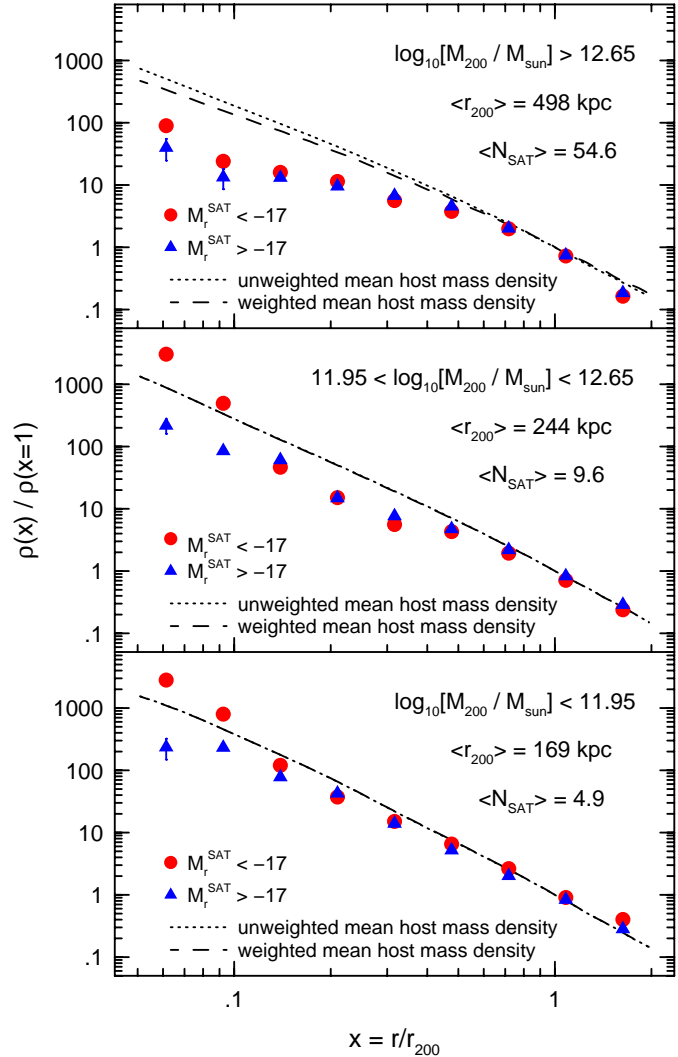}
\vskip -2.5cm
\caption{Normalized satellite number density profiles 
and host mean mass density profiles as a function of host virial
mass.
Satellites are divided in luminosity as in Figure~2.
Unweighted means and means weighted 
by the number of satellites per host are shown
for the host mass density.
{\it Top:} most massive hosts 
($M_{200} > 4.5\times 10^{12} M_\odot$; 92 hosts).
{\it Middle:} hosts with $8.9\times 10^{11} M_\odot \le M_{200} 
\le 4.5\times 10^{12} M_\odot$ (472 hosts).
{\it Bottom:} least massive hosts ($M_{200} < 8.9\times 10^{11}
M_\odot$; 461 hosts).
}
\label{fig:profiles3}
\end{figure}

\section{Summary and Discussion}

The analysis above shows that the number density profiles
for Illustris-1 satellite galaxies differ substantially from the
number density profiles of  
similar systems obtained using SAMs.
The most striking difference
is that on scales
$\lesssim 0.4~ r_{200}$ the complete sample of
Illustris-1 satellites does not trace the surrounding mass density of the hosts.
The number density profile of the brightest satellites ($M_r < -17$,
comparable to SNLWC's satellites) cannot be fitted by an NFW profile
due to a steep increase in the slope of the number density profile
for host-satellite separations
$r \lesssim 0.1~ r_{200}$. 
The number density profile of the
faintest satellites ($M_r > -17$) is well-fitted by an NFW profile, but the concentration
of the best-fitting NFW profile is much
lower than the concentration of the mean host mass density. That is,
the distribution of faint satellites is significantly less concentrated than
is the mass surrounding the hosts.
Weighting the mean host mass density by the number of satellites per
host does not affect this conclusion.
In addition, the shapes of the satellite number density profiles for Illustris-1
satellites show a clear dependence on host luminosity and
virial mass that was not seen 
by SNLWC.

A key difference between the Illustris-1 satellites and
SNLWC's satellites
is the methodology by which the luminous galaxy catalogs were
created; i.e., a hydrodynamical simulation vs.\ a SAM.  
Because the luminous galaxy catalogs resulted from different 
techniques, some differences should be expected. In particular, significant
differences might be expected to manifest on small host-satellite separations.
This is due to the different ways in which satellites, orbiting near the 
host galaxies, are treated.  Satellites in hydrodynamical simulations are followed
by the simulation directly. They are tidally-stripped, disrupted, and
merge with their hosts on a time scale that  relies on
direct integration of the equations of motion.  In SAMs, once satellite
galaxies orbiting close to their hosts are stripped of so
much dark matter that the mass of their subhalo drops below the resolution limit,
the satellite can only be identified 
by the position and velocity of a single particle. That single particle corresponds to 
the most bound particle at the last time the satellite's
subhalo could be resolved. Once this point is reached,
the SAM merges the satellite with the host on a particular
time scale.  In the case of the Croton et al.\ (2006) SAM used by 
SNLWC, stripped satellites 
were merged with their host on a time scale set by dynamical friction.
The steep upturn in the number density profile for Illustris-1 satellites with
$M_r < -17$ on scales $\lesssim 0.1~ r_{200}$, which was not found by 
SNLWC for similar MS satellites, may indicate that 
the Croton et al.\ (2006) SAM merged satellites with their hosts faster than 
would have occurred if the MS had incorporated numerical 
hydrodynamics.

The difference in force resolution between Illustris-1 and the MS is 
another factor that could contribute to the differences between the
density profiles found here and those found by SNLWC.
However, in their study of satellite galaxies in both the MS and the MS-II (which had a 
force softening five times smaller than the MS and only a factor of two
larger than Illustris-1), Wang et al.\ (2014) did not find a 
steep increase in the satellite number density profile for small host-satellite
separations in the MS-II. Wang et al.\ (2014) concluded the 
differences between the number density profiles of observed satellites and those
obtained from SAMs could be attributed to environmental effects in the SAMs
being too efficient.

The steep increase in the number density profile 
exhibited by bright ($M_r < -17$) Illustris-1 satellites at small host-satellite
separations is in qualitative agreement with the 
observational results of Tal et al.\ (2012), 
Watson et al.\ (2012), and 
Piscionere et al.\ (2015), all of which concluded that the
number density profile of bright satellites increases on small 
scales. In their study of the satellites of LRGs (i.e., host 
galaxies that, on average, are much brighter than the Illustris-1 hosts), Tal
et al.\ (2012) concluded that the steep increase in the satellite number 
density profile could be explained by a corresponding steep increase in the
luminous mass density of the hosts on small scales.  In particular, Tal et al.\
(2012) concluded that, on scales $r \lesssim 25$~kpc, the baryonic mass of the 
host galaxies accounts for $\gtrsim 50$\% of the total mass of the hosts.
For the Illustris-1 hosts, no steep increase in the mass density 
profile occurs on small scales.  Rather, the
total mass density profile (i.e., baryonic plus dark matter)
of the Illustris-1 hosts is well-fitted by an NFW profile.
Within a radius
of 25~kpc (comparable to 10\% of the median virial radius of the host
sample, $r_{200}^{\rm med} = 205$~kpc), the baryonic mass fraction of the 
Illustris-1 hosts is 
$\sim 26$\%, rather than the $\gtrsim 50$\% obtained by Tal et al.\ (2012)
for their LRG hosts.  Given that the host galaxies in Tal et al.\ (2012) 
are much more massive than the vast majority of the Illustris-1 hosts,
this difference in the small-scale baryonic mass fraction may not be significant.
However, the present results do show that a steep increase in the satellite 
number density profile at small host-satellite separations does not 
require a corresponding 
steep increase in the host mass since, in the case of Illustris-1
satellites, the satellites simply do not trace the host mass on small scales.

The results presented here are, of course, dependent upon a particular 
simulation and the distribution of satellite galaxies for small host-satellite 
separations may be sensitive to approximations to the detailed
gas physics that are adopted in the simulation.  That in mind, it will be especially
interesting to compare the present results to a similar analysis of the IllustrisTNG 
simulations (e.g., Weinberger et al.\ 2017; Nelson et al.\ 2018;
Pillepich et al.\ 2018), 
which adopted different models for the growth of supermassive black holes,
galactic winds driven by stellar feedback, and 
AGN feedback.  
These modifications resulted in significant improvements of IllustrisTNG over 
the original Illustris Project, including a
stellar mass function for the simulated galaxies that agrees with observations
and a clear red-blue galaxy color bimodality that was
not seen in the original Illustris Project.
Other high resolution hydrodynamical simulations to which it would be interesting
to compare the present results include the EAGLE simulations (e.g., 
Schaye et al.\ 2014; Crain et al.\ 2015),
the Magneticum simulations (Hirschmann et al.\ 2014; Teklu et al.\ 2015), and the
MassiveBlack-II simulation (Khandai et al.\ 2014).

\section*{Acknowledgments}

Insightful conversations with Patrick Koh  and Masaya Yamamoto are gratefully
acknowledged.



\end{document}